\documentclass{svjour3}                     
\smartqed  
\usepackage{graphicx}
\usepackage{multirow}
\begin{document}

\title{Three body systems with strangeness and exotic systems \footnote{Presented at the 21st European Conference on Few-Body Problems in Physics, Salamanca, Spain, 30 August - 3 September 2010}
}


\author{E. Oset         \and
        A. Martinez Torres \and
	K. P. Khemchandani \and
	L. Roca 
}


\institute{E. Oset \at
              Departamento de Fisica Teorica and IFIC, Centro Mixto Universidad de Valencia-CSIC, Institutos de Investigacion
de Paterna, Aptdo. 22085, 46071 Valencia, Spain \\
              Tel.: +34-963543525\\
              Fax:  +34-963543488\\
              \email{oset@ific.uv.es}           
           \and
        A. Martinez Torres \at
                  Yukawa Institute for Theoretical Physics, Kyoto University, Kyoto 606-8502, Japan
		  \and
		  K.P. Khemchandani \at
	Research Center for Nuclear Physics (RCNP), Osaka University, Ibaraki, Osaka 567-0047, Japan
	\and
	L. Roca \at
	Departamento de Fisica. Universidad de Murcia. E-30071, Murcia. Spain
}

\date{Received: date / Accepted: date}

\maketitle

\begin{abstract}
We report on four $\Sigma$'s and three $\Lambda$'s, in the 1500 - 1800 MeV region, as two meson - one baryon S-wave $(1/2)^+$ resonances found by solving the Faddeev equations in the coupled channel approach, which can be associated  to the existing $S$ = -1, $J^P= 1/2^+$ low lying baryon resonances. On the other hand we also report on a new, hidden strangeness $N^*$ state, mostly made of $K \bar{K} N$, with mass around 1920 MeV, which we think could be responsible for the peak seen in the $\gamma p \to K^+ \Lambda$ around this energy. Finally we address a very novel topic in which we show how few body systems of several $\rho$ mesons can be produced, with their spins aligned up to J=6, and how these states found theoretically can be associated to several known mesons with spins J=2,3,4,5,6.

\end{abstract}

\section{Introduction}
\label{intro}
Our understanding of baryon resonances is undergoing continuous change. From the classical picture of the baryons made out of three constituent quarks, passing through attempts to represent some of them in terms of pentaquarks, to the more recent description of some of them in terms of meson baryon molecules. In this sense, the low lying $J^P= 1/2^-$ resonances, $\Lambda(1405)$, $\Lambda(1670)$ $\dots$, can be represented as dynamically generated states from the meson baryon interaction and are relatively well understood within the unitary chiral models \cite{Kaiser:1995eg,angels,ollerulf,Nieves:1999bx,jido,carmen,hyodo,borasoy}. The low lying $J^P= 1/2^+$ domain remains far less understood, both experimentally and theoretically. For instance, quark models seem to face difficulties in reproducing properties of the resonances in this sector \cite{riska}. The neat reproduction of the low lying $1/2^-$ states in the $S$-wave meson-baryon interaction, using chiral dynamics, suggests that the addition of a pseudoscalar meson in S-wave could lead to an important component of the structure of the $1/2^+$ resonances.  Chiral dynamics has been used earlier in the context of the three nucleon problems, e.g., in \cite{epelbaum}. We present in this talk the study done in \cite{MartinezTorres:2007sr} of two meson - one baryon systems where chiral dynamics is applied to solve the Faddeev equations. As shall be described below, our calculations for the total $S$ = -1 reveal peaks in the  $\pi \bar{K} N$ system and its coupled channels which we identify with the resonances $\Sigma(1770)$, $\Sigma(1660)$, $\Sigma(1620)$, $\Sigma(1560)$, $\Lambda(1810)$ and $\Lambda(1600)$.

\section{The formalism for three body systems}
We start by taking all combinations of a pseudoscalar meson of the $0^-$ SU(3) octet and a baryon of the $1/2^+$ octet which couple to $S=-1$ with any charge. For some quantum numbers, the interaction of this two body system is strongly attractive and responsible for the generation of the two $\Lambda(1405)$ states \cite{jido} and other $S$ = -1 resonances. We shall assume that this two body system formed by $\bar{K}N$ and coupled channels remains highly correlated when a third particle is added, in the present case a pion. Altogether, we get twenty-two coupled channels for the net charge zero configuration: $\pi^0 K^- p$, $\pi^0\bar{K}^0 n$, $\pi^0\pi^0\Sigma^0$, $\pi^0\pi^+\Sigma^-$, $\pi^0\pi^-\Sigma^+$, $\pi^0\pi^0\Lambda$, $\pi^0\eta\Sigma^0$, $\pi^0\eta\Lambda$, $\pi^0 K^+\Xi^-$, $\pi^0 K^0\Xi^0$, $\pi^+ K^- n$, $\pi^+\pi^0\Sigma^-$, $\pi^+\pi^-\Sigma^0$, $\pi^+\pi^-\Lambda$, $\pi^+\eta\Sigma^-$, $\pi^+ K^0\Xi^-$, $\pi^-\bar{K}^0 p$, $\pi^-\pi^0\Sigma^+$, $\pi^-\pi^+\Sigma^0$, $\pi^-\pi^+\Lambda$, 
$\pi^-\eta\Sigma^+$, $\pi^- K^+ \Xi^0$. We assume the correlated pair to have a certain invariant mass, $\sqrt{s_{23}}$, and the three body $T$-matrix is evaluated as a function of  this mass and the total energy of the three body system. At the end we look for the value of $|T|^2$ as a function of these two variables and find peaks at certain values of these two variables, which indicate the mass of the resonances and how a pair of particles is correlated.  

The input required to solve the Faddeev equations, i.e., the two body $t$-matrices for the meson-meson and meson-baryon interactions has been calculated by taking the lowest order chiral Lagrangian following
\cite{npa,angels,bennhold,Inoue} and using the dimensional regularization of the loops as done in
\cite{ollerulf,bennhold}, where a good reproduction of scattering amplitudes and resonance properties was found. Instead, a cut off could also be used to regularize the loops as shown in \cite{angels,ollerulf}. Improvements introducing higher order Lagrangians have been done recently, including a theoretical error analysis \cite{boraulf} which allows one to see that the results with the lowest order Lagrangian fit perfectly within the theoretical allowed bands.

A shared feature of the recent unitary chiral dynamical calculations is the on-shell factorization of the potential and the $t$-matrix in the Bethe-Salpeter equation \cite{npa,angels,ollerulf,Nieves:1999bx,carmen,hyodo,borasoy}, which is
justified by the use of the N/D method and dispersion relations \cite{nsd,ollerulf}. Alternatively, one can see that the off-shell contributions can be reabsorbed into renormalization of the lower order terms \cite{npa,angels}. We develop here a similar approach for the Faddeev equations.

The full three-body $T$-matrix can be written as a sum of the auxiliary $T$-matrices $T^1$, $T^2$ and $T^3$ \cite{Faddeev}
\begin{equation}
T=T^1+T^2+T^3
\end{equation}
where $T^i$, $i=1$, $2$, $3$, are the normal Faddeev partitions, which include all the possible interactions contributing to the three-body $T$-matrix with the particle $i$ being a spectator in the last interaction.
The Faddeev partitions satisfy the equations
\begin{equation}\label{eq:Tiorig}
T^i=t^i\delta^3(\vec{k}^{\,\prime}_i-\vec{k}_i)+ t^i g^{ij}T^j + t^i g^{ik}T^k ,
\end{equation}
where $\vec{k}_i$ ($\vec{k}^{\,\prime}_i$) is the initial (final) momentum of the ith particle in the global center of mass system, $t^i$ is the two-body $t$-matrix for the interaction of the 
pair $(jk)$ and $g^{ij}$ is the three-body propagator or Green's function, with $j \neq k \neq i$ = 1, 2, 3

\begin{figure}[ht]
\includegraphics[width=0.5\textwidth] {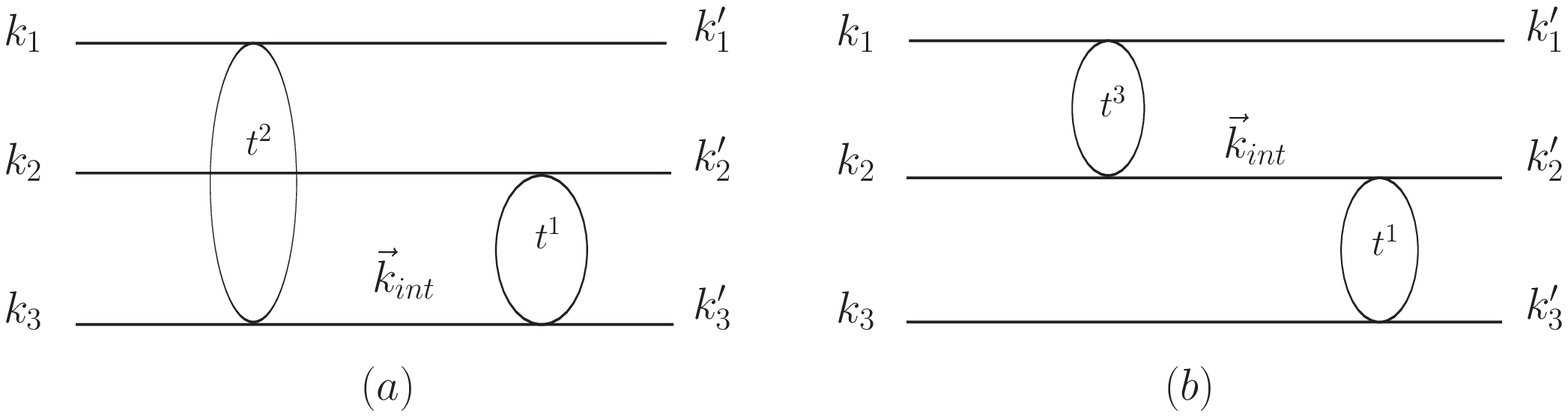}
\caption{\label{fig1} Diagrammatic representation of the terms (a) $t^1 g^{12} t^2$
(b) $t^1 g^{(13)} t^3$.}
\end{figure}

\begin{figure}[ht]
\includegraphics[scale=0.6] {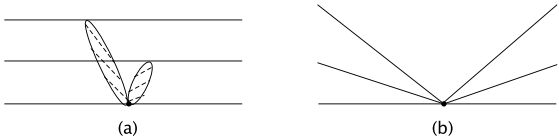}
\caption{\label{fig2} The origin of the three body forces (a) due to cancellation of the propagator in Fig.\ref{fig1}(a) with the off-shell part of the chiral amplitude, (b) at the tree level.}
\end{figure}

The first two terms of the diagrammatic expansion of the Faddeev equations, for the case $i$=1, are represented diagrammatically in Fig.\ref{fig1},
where the $t$-matrices are required to be off-shell. However, the chiral amplitudes, which we use,
can be split into an ``on-shell'' part (obtained when the only propagating particle of the diagrams, labeled with $\vec{k}_{int}$ in Fig.\ref{fig1}, is placed on-shell, meaning that $q^2$ is replaced by $m^2$ in the amplitudes), which depends only on the c.m energy of the interacting pair, and an off-shell part proportional to the inverse of the propagator of the off-shell particle. This term would cancel the particle propagator, ($q^2-m^2$), for example that of the 3rd particle in the Fig.\ref{fig1}a) resulting into a three body force (Fig.\ref{fig2}a). In addition to this,
three body forces also stem directly from the chiral Lagrangians \cite{Felipe} (Fig.\ref{fig2}b).

We find that the sum of the off-shell parts of all the six $t^i g^{ij} t^j$ terms, together with the contribution from Fig.\ref{fig2}(b) cancels exactly when the SU(3) limit is considered. Details of the analytical proof can be seen in the appendices of \cite{MartinezTorres:2007rz,MartinezTorres:2008gy}. Hence, only the on-shell part of the two body (chiral) $t$-matrices is needed in the evaluations. This is one of the important findings of these works because one of the standing problems of the Faddeev equations is that the use of different potentials which give rise to the same on shell scattering amplitudes give rise to different results when used to study three body system with the Faddeev equations. The different, unphysical, off shell amplitudes of the different potentials are responsible for it. The use of chiral dynamics in the context of the Faddeev equations has then served to show that the results do not depend on these unphysical amplitudes and only the on shell amplitudes are needed as input. In this sense, since these amplitudes can be obtained from experiment, it is suggested in \cite{MartinezTorres:2008kh} to use these experimental amplitudes, and sensible results are obtained in the study of the $\pi \pi N $ system and coupled channels.

\begin{figure}[ht]
\includegraphics[scale=0.40] {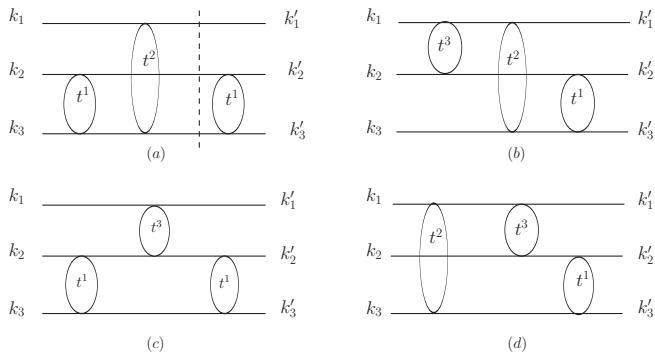}
\caption{\label{fig3} Different diagrams involving three pair interactions corresponding  to the $T^1$ partition.}
\end{figure}

The first term with a non trivial structure, from the point of view of the on-shell factorization of the $t$-matrices in the Faddeev equations, is the one involving three successive pair interactions, where a loop function of three particle propagators appears for the first time. We show the diagrams with such a structure for the $T^1$ partition in Fig.\ref{fig3}(a-d).

 The strategy followed in the former works is that the terms with two, three, and four interactions are evaluated exactly. Then it is observed that the ratio of the four to three body interaction terms is about the same as that of the three body to two body. Once this is realized, the coupled integral equations are converted into algebraic equations, which renders the technical work feasible in spite of the many coupled channels used.

There are two independent variables in the formalism $\sqrt{s}$, $\sqrt{s_{23}}$, as a function of which we plot the squared $T^*_R$-matrix ($T^*_R = \sum\limits_{ij} ( T_R^{ij} - t^i g^{ij}t^j$ ) ), where $T_R^{ij}$ sum all terms with the first two interactions having $i,j$ as spectators.

\begin{figure}[ht]
\includegraphics[width=0.6\textwidth]{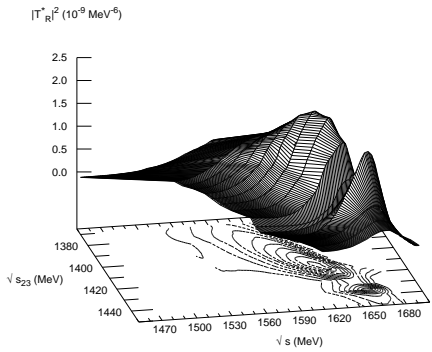}
\caption{\label{fig4} Two $\Sigma$ resonances in the  $\pi \pi \Sigma$ amplitude in $I$ = 1, $I_\pi$ = 2 configuration.}
\end{figure}

We now report on the four isospin I=1 states found in our study. In Fig.\ref{fig4}, we show a plot of the squared $T^*_R$-matrix and its projection, for $\pi \pi \Sigma \,\rightarrow \, \pi \pi \Sigma$ in the total isospin $I$ = 1 configuration obtained by keeping the two pions in isospin $I_\pi$ = 2. We see two peaks; one at $\sqrt{s}$ = 1656 MeV with the full width at half maximum $\sim$ 30 MeV and another at $\sqrt{s}$ = 1630 MeV with $\Gamma$ = 39 MeV. We identify the peak at $\sqrt{s}$ = 1656 MeV with the well established $\Sigma(1660 - i100/2)$  \cite{pdg} as a resonance in the $\pi \pi \Sigma$ system, which is a new finding.

A detailed description of all the states that appear in this sector can be seen in \cite{MartinezTorres:2007sr}. Here we summarize the results in Table \ref{table1}.

\begin{table}
\centering
\begin{tabular}{cccc}
\hline
&$\Gamma$ (PDG)&Peak position& $\Gamma$ (this work)\\
&(MeV)&(this work, MeV)& (MeV)\\
\hline
Isospin=1&&&\\
\hline
$\Sigma(1560)$&10-100&1590&70\\
$\Sigma(1620)$&10-100&1630&39\\
$\Sigma(1660)$&40-200&1656&30\\
$\Sigma(1770)$&60-100&1790&24\\
\hline
Isospin=0&&&\\
\hline
$\Lambda(1600)$&50-250&1568,1700&60-136\\
$\Lambda(1810)$&50-250&1740&20\\
\hline

\end{tabular}
\caption{$\Sigma$ and $\Lambda$ states obtained from the interaction of two mesons and one baryon.}\label{table1}
\end{table}

  In the S=0 sector we also find several resonances, which are summarized in Table \ref{table2}. Since the talk is mostly about strangeness, we only want to pay attention to the $N^*$ state around 1920 MeV, which is mostly $N K \bar{K}$. This state was first predicted in \cite{Jido:2008kp} using variational methods and corroborated in \cite{MartinezTorres:2008kh} using coupled channels Faddeev equations. As in \cite{Jido:2008kp}, we find that the $ K \bar{K}$ pair is built mostly around the $f_0(980)$, but it also has a similar strength around the $a_0(980)$, both of which appear basically as a $ K \bar{K}$ molecule in the chiral unitary approach.
  
\begin{table}
\centering
\begin{tabular}{c|ccc|ccc}
\hline\hline
$I(J^P)$&\multicolumn{3}{c}{Theory}&\multicolumn{3}{c}{PDG data}\\
\hline
&channels&mass&width&name&mass&width\\
&&(MeV)&(MeV)&&(MeV)&(MeV)\\
\hline
$1/2(1/2^+)$&only $\pi\pi N$&1704&375&$N^*(1710)$&1680-1740&90-500\\
& $\pi\pi N$, $\pi K\Sigma$, $\pi K\Lambda$, $\pi\eta N$&$\sim$ no change&$\sim$ no change&&&\\
\hline
$1/2(1/2^+)$&only $\pi\pi N$&2100&250&$N^*(2100)$&1885-2270&80-400\\
& $\pi\pi N$, $\pi K\Sigma$, $\pi K\Lambda$, $\pi\eta N$&2080&54&&&\\
\hline
$3/2(1/2^+)$&$\pi\pi N$, $\pi K\Sigma$, $\pi K\Lambda$, $\pi\eta N$&2126&42&$\Delta(1910)$&1870-2152&190-270\\
\hline
$1/2(1/2^+)$&$N\pi\pi $, $N\pi\eta $, $NK\bar K$&1924&20&$N^*(?)$&?&?\\
\hline\hline
\end{tabular}
\caption{$N^{*}$ and $\Delta$ states obtained from the interaction of two mesons and one baryon.}\label{table2}
\end{table}
  
  This state is very interesting and it has been suggested in  \cite{conulf} that it could be responsible for the peak around 1920 MeV of the $\gamma p \to K^+ \Lambda$ reaction \cite{saphir,jefflab,mizuki}. It was suggested that the spin of the resonance could be found performing polarization measurements which are the state of the art presently.
  
In case the $J^P=1/2^+$ assignment was correct, an easy test can be carried out to rule out the $3/2^+$ state. 
The experiment consists in performing the 
$\gamma p \to K^+ \Lambda$ reaction with a
circularly polarized photon with helicity~1, thus $S_z=1$ with the $z$-axis
defined along the photon direction, together with a polarized proton of the
target with $S_z=1/2$ along the same direction. With this set up, the total spin
has $S^{tot}_z=3/2$. Since $L_z$ is zero with that choice of the $z$ direction,
then $J^{tot}_z=3/2$ and $J$ must be equal or bigger than $3/2$. Should the
resonant state be $J^P=1/2^+$, the peak signal would disappear for this polarization
selection, while it would remain if the resonance was a $J^P=3/2^+$ state. 
Thus, the disappearance of the signal with this polarization set up would rule out the 
$J^P=3/2^+$ assignment.

  Such type of polarization set ups have been done and are common in facilities
like ELSA at Bonn, MAMI B at Mainz or CEBAF at Jefferson Lab, where spin-3/2
and 1/2 cross sections, which play a crucial role
in the GDH sum rule, see e.g. Ref.~ \cite{Drechsel:1995az}, were measured in the
two-pion photoproduction \cite{Ahrens:2001qt,Ahrens:2007zzj} reaction.
The theoretical analysis of \cite{Nacher:2001yr} shows indeed that the
separation of the amplitudes in the spin channels provides information on the
resonances excited in the reaction.

  In \cite{conulf} it was also shown that the presence of this resonance would lead to an enhanced $K \bar{K}$ invariant mass distribution close to the 
$K \bar{K}$ threshold, and also it would produce an enhanced cross section  in the  $\gamma p \to K^+ K^- p$ close to threshold of the reaction.  These features are now being tested at Spring8/Osaka \cite{nakaprivate}.

\section{Multirho states}   
To finalize this selection of topics, we would like to address a recent study that finds some unexpected few body systems. These are states with several bound $\rho$ mesons, up to six. 
  In the PDG~\cite{pdg}
there are intriguing mesons with large spin, of the $\rho$ and $f_0$
type, whose quantum numbers match systems made with 3, 4, 5 and 6 $\rho$ mesons with their spins aligned. 
These are the $\rho_3(1690)$ ($3^{--}$),   $f_4(2050)$ ($4^{++}$),
 $\rho_5(2350)$ ($5^{--}$) and   $f_6(2510)$ ($6^{++}$)
resonances.  The idea of having these states as multirho states stems from the findings of \cite{Molina:2008jw} where the $\rho\rho$ interaction was studied using a unitary coupled channel approach with the input from the hidden gauge Lagrangians \cite{hidden1,hidden2,hidden3,hidden4}. In the work \cite{Molina:2008jw}
it was found that the interaction of two $\rho(770)$ mesons in isospin
$I=0$ and spin $S=2$ was strong enough to bind the $\rho\rho$ system
into the $f_2(1270)$ ($J^{PC}=2^{++}$) resonance.  The nature of this
resonance as a   $\rho(770)\rho(770)$ molecule has passed the tests of
radiative decay into $\gamma \gamma$ \cite{junko}, the decay of $J/\Psi$
into $\omega (\phi)$ and $f_2(1270)$ (together with other resonances
generated in \cite{gengvec})  \cite{daizou}, and $J/\Psi$ into $\gamma$
and  $f_2(1270)$ (and the other resonances of \cite{gengvec})
\cite{hanhart}.

\begin{figure}[!t]
\begin{center}
\includegraphics[width=0.7\textwidth]{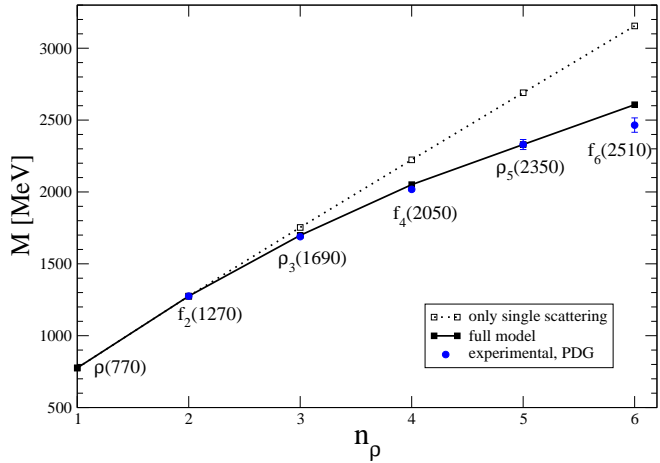}
\caption{Masses of the dynamically generated states as a function of the
number of constituent $\rho(770)$ mesons, $n_\rho$. Only single
scattering contribution (dotted line); full model (solid line);
experimental values from the PDG\cite{pdg}, (circles).
}
\label{fig:Mvsn}
\end{center}
\end{figure}

   The realization of this strong attraction in J=2, suggest that other $\rho$ mesons could cluster, always with their spin aligned such that all pairs would have J=2.  This is what has been done in \cite{Roca:2010tf}. One studies the interaction of a $\rho$ with the $f_2(1270)$, using the fixed center approximation to the Faddeev equations, and the scattering matrix leads to a clear bump that we associate to the $\rho_3(1690)$ ($3^{--}$). After this, one studies the interaction of two $f_2(1270)$, where the input is the scattering amplitude of $\rho$ with the $f_2(1270)$, studied before. A peak is also found in the scattering matrix, which we associate to the $f_4(2050)$ ($4^{++}$). Further iterations including every time one extra $\rho$ allow us to find also peaks in amplitudes that we associate to the $\rho_5(2350)$ ($5^{--}$) and   $f_6(2510)$ ($6^{++}$) states.  The agreement found with experiment, at the cost of no free parameters, is excellent, as one can see in Fig. \ref{fig:Mvsn}.
   
     It would be interesting to apply these ideas to other systems where one of the $\rho$ mesons is replaced by a $K^*$ in order to see if one obtains the several high spin $K^*$ stated of \cite{pdg}. Work along this direction  has just appeared \cite{junkonew} and the predictions are equally successful.

\section{Summary}
We conclude the discussion by emphasizing that all the low lying ${1/2}^+$ $\Sigma$ and $\Lambda$ resonances in the $PDG$ \cite{pdg}, up to the 1800 MeV energy region, get dynamically generated as two meson-one baryon states
in these calculations. In addition, we predict the quantum numbers of the $\Sigma(1560)$ and also find evidence 
for a  ${1/2}^+$ $\Sigma$ resonance at $\sim$ 1620 MeV. 

  On the other hand we also reported on a hidden strangeness three body system, which leads to a new $N^*$ state with mass around 1920 MeV and which we think is responsible for the peak seen in the $\gamma p \to K^+ \Lambda$ reaction. We suggest to perform experiments on polarization and also the study of the 
  $\gamma p \to K^+ K^- p$ reaction close to threshold in order to test the predictions tied to the existence of this resonance below the  $K^+ K^- p$ threshold.
  
    Finally we reported on very novel few body systems, states made of many $\rho$ mesons with their spins aligned. We could find that these states are bound, although they have a width from decay to pions. Yet, up to J=6, the widths are still on the measurable range such that we could associate the states found to several existing mesons with spin up to J=6.
    
    As a general remark, the combination of the chiral  unitary techniques with few body systems is proving to be a very fruitful field to which we can only encourage our colleagues to jump to.

\section{Acknowledgements}
 This work is partly
supported by DGICYT contract number FIS2006-03438
and the Generalitat Valenciana in the program Prometeo.
We acknowledge the support of the European
Community-Research Infrastructure Integrating Activity
Study of Strongly Interacting Matter (acronym
HadronPhysics2, Grant Agreement n. 227431) under the
Seventh Framework Programme of EU.


\begin{thebibliography}{99}


\bibitem{Kaiser:1995eg}
  N.~Kaiser, P.~B.~Siegel and W.~Weise,
  Nucl.\ Phys.\  A {\bf 594}, 325 (1995)
.
  
\bibitem{angels}
  E.~Oset and A.~Ramos,
  Nucl.\ Phys.\  A {\bf 635}, 99 (1998).
   
\bibitem{ollerulf}
  J.~A.~Oller and U.~G.~Meissner,
  Phys.\ Lett.\  B {\bf 500}, 263 (2001)
  




\bibitem{Nieves:1999bx}
  J.~Nieves and E.~Ruiz Arriola,
  Nucl.\ Phys.\  A {\bf 679}, 57 (2000)
  
\bibitem{jido}
  D.~Jido {\it et. al.},
  Nucl.\ Phys.\  A {\bf 725}, 181 (2003)

\bibitem{carmen}
  C.~Garcia-Recio, M.~F.~M.~Lutz and J.~Nieves,
  Phys.\ Lett.\  B {\bf 582}, 49 (2004)

\bibitem{hyodo}
  T.~Hyodo, S.~I.~Nam, D.~Jido and A.~Hosaka,
  Phys.\ Rev.\  C {\bf 68}, 018201 (2003)
  
\bibitem{borasoy}
  B.~Borasoy, R.~Nissler and W.~Weise,
  Eur.\ Phys.\ J.\  A {\bf 25}, 79 (2005)
  
\bibitem{riska}
  L.~Y.~Glozman and D.~O.~Riska,
  Phys.\ Rept.\  {\bf 268}, 263 (1996)
.

%
  
\bibitem{epelbaum}
  E.~Epelbaum {\it et. al.},
  Phys.\ Rev.\ Lett.\  {\bf 86}, 4787 (2001)
  
\bibitem{MartinezTorres:2007sr}
  A.~Martinez Torres, K.~P.~Khemchandani and E.~Oset,
  Phys.\ Rev.\  C {\bf 77}, 042203 (2008)
  
\bibitem{npa}
  J.~A.~Oller and E.~Oset,
  Nucl.\ Phys.\  A {\bf 620}, 438 (1997)
  [Erratum-ibid.\  A {\bf 652}, 407 (1999)].
  

\bibitem{bennhold}
  E.~Oset, A.~Ramos and C.~Bennhold,
  Phys.\ Lett.\  B {\bf 527}, 99 (2002)
  [Erratum-ibid.\  B {\bf 530}, 260 (2002)].

\bibitem{Inoue}
  T.~Inoue, E.~Oset and M.~J.~Vicente Vacas,
  Phys.\ Rev.\  C {\bf 65}, 035204 (2002)


  
\bibitem{boraulf}
  B.~Borasoy, U.~G.~Meissner and R.~Nissler,
  Phys.\ Rev.\  C {\bf 74}, 055201 (2006)
  
  
\bibitem{nsd}
  J.~A.~Oller and E.~Oset,
  Phys.\ Rev.\  D {\bf 60}, 074023 (1999).



\bibitem{Faddeev}
  L.~D.~Faddeev,
  Sov.\ Phys.\ JETP {\bf 12}, 1014 (1961)
  [Zh.\ Eksp.\ Teor.\ Fiz.\  {\bf 39}, 1459 (1960)].


\bibitem{Felipe}
 F.~J.~Llanes-Estrada, E.~Oset and V.~Mateu,
 Phys.\ Rev.\  C {\bf 69}, 055203 (2004).
 
 
\bibitem{MartinezTorres:2007rz}
  A.~Martinez Torres, K.~P.~Khemchandani and E.~Oset,
  Eur.\ Phys.\ J.\  A {\bf 36}, 211 (2008)
  [arXiv:0712.1938 [nucl-th]].

\bibitem{MartinezTorres:2008gy}
  A.~Martinez Torres, K.~P.~Khemchandani, L.~S.~Geng, M.~Napsuciale and E.~Oset,
  Phys.\ Rev.\  D {\bf 78}, 074031 (2008)
  [arXiv:0801.3635 [nucl-th]].




\bibitem{MartinezTorres:2008kh}
  A.~Martinez Torres, K.~P.~Khemchandani and E.~Oset,
  Phys.\ Rev.\  C {\bf 79}, 065207 (2009)

\bibitem{pdg}
  C.~Amsler {\it et al.}  [Particle Data Group],
  Phys.\ Lett.\  B {\bf 667}, 1 (2008).

  
%
\bibitem{Jido:2008kp}
  D.~Jido and Y.~Kanada-En'yo,
  Phys.\ Rev.\  C {\bf 78}, 035203 (2008)

\bibitem{conulf}
  A.~Martinez Torres, K.~P.~Khemchandani, U.~G.~Meissner and E.~Oset,
  Eur.\ Phys.\ J.\  A {\bf 41}, 361 (2009)
  
\bibitem{saphir}
 K.~H.~Glander {\it et al.},
 Eur.\ Phys.\ J.\  A {\bf 19}, 251 (2004)

\bibitem{jefflab}
 R.~Bradford {\it et al.}  [CLAS Collaboration],
 Phys.\ Rev.\  C {\bf 73}, 035202 (2006)

\bibitem{mizuki}
 M.~Sumihama {\it et al.}  [LEPS Collaboration],
 Phys.\ Rev.\  C {\bf 73}, 035214 (2006)
 
 
\bibitem{Drechsel:1995az}
 D.~Drechsel,
 Prog.\ Part.\ Nucl.\ Phys.\  {\bf 34}, 181 (1995)



\bibitem{Ahrens:2001qt}
 J.~Ahrens {\it et al.}  [GDH Collaboration and A2 Collaboration],
 Phys.\ Rev.\ Lett.\  {\bf 87}, 022003 (2001)


\bibitem{Ahrens:2007zzj}
 J.~Ahrens {\it et al.}  [GDH and A2 Collaborations],
 Eur.\ Phys.\ J.\  A {\bf 34}, 11 (2007).

\bibitem{Nacher:2001yr}
 J.~C.~Nacher and E.~Oset,
 Nucl.\ Phys.\  A {\bf 697}, 372 (2002)
 
\bibitem{nakaprivate} T. Nakano, private comunication. 

\bibitem{Molina:2008jw}
  R.~Molina, D.~Nicmorus and E.~Oset,
  Phys.\ Rev.\  D {\bf 78}, 114018 (2008)


  
\bibitem{hidden1}
  M.~Bando, T.~Kugo, S.~Uehara, K.~Yamawaki and T.~Yanagida,
  Phys.\ Rev.\ Lett.\  {\bf 54} (1985) 1215.


\bibitem{hidden2}
  M.~Bando, T.~Kugo and K.~Yamawaki,
  Phys.\ Rept.\  {\bf 164} (1988) 217.

\bibitem{hidden3}
  M.~Harada and K.~Yamawaki,
  Phys.\ Rept.\  {\bf 381}, 1 (2003).
  
\bibitem{hidden4}
  U.~G.~Meissner,
  Phys.\ Rept.\  {\bf 161}, 213 (1988).

\bibitem{junko}
  H.~Nagahiro, J.~Yamagata-Sekihara, E.~Oset and S.~Hirenzaki,
  Phys.\ Rev.\  D {\bf 79}, 114023 (2009).
  
\bibitem{gengvec}
  L.~S.~Geng and E.~Oset,
  Phys.\ Rev.\  D {\bf 79}, 074009 (2009).
  
\bibitem{daizou}
  A.~Martinez Torres, L.~S.~Geng, L.~R.~Dai, B.~X.~Sun, E.~Oset and B.~S.~Zou,
  Phys.\ Lett.\  B {\bf 680}, 310 (2009).
  

  
\bibitem{hanhart}
  L.~S.~Geng, F.~K.~Guo, C.~Hanhart, R.~Molina, E.~Oset and B.~S.~Zou,
  Eur.\ Phys.\ J.\  A {\bf 44}, 305 (2010)


  
\bibitem{Roca:2010tf}
  L.~Roca and E.~Oset,
  Phys.\ Rev.\  D {\bf 82}, 054013 (2010)
  
   
\bibitem{junkonew}
  J.~Yamagata-Sekihara, L.~Roca and E.~Oset,
  arXiv:1010.0525 [hep-ph].






\end{thebibliography}

\end{document}